\documentclass{article}[12pt]

\usepackage{graphics,color}
\usepackage{graphics}
\usepackage{epsfig}
\usepackage{amsmath}
\usepackage{amssymb}

\begin{document}
\newcommand {\CO} {COMPASS }
\newcommand {\PO} {POLAR }
\newcommand\arcdeg{\mbox{$^\circ$}}%

\title{\CO: An Instrument for Measuring the Polarization of the CMB on
Intermediate Angular Scales}

\author{Philip C. Farese (Princeton), 
Giorgio Dall'Oglio (Rome), 
Josh Gundersen (U Miami),\\
Brian Keating (CalTech), 
Slade Klawikowski (UW), 
Lloyd Knox (UC Davis), \\
Alan Levy (UCSB),
Philip Lubin (UCSB) 
Chris O'Dell (UMass), 
Alan Peel (UC Davis), \\ 
Lucio Piccirillo (Cardiff), 
John Ruhl (CWRU), 
Peter Timbie (UW)}
\maketitle

\begin{abstract}
\CO is an on-axis 2.6 meter telescope coupled to a correlation
polarimeter.  The entire instrument was built specifically for CMB
polarization studies.  Careful attention was given to receiver and
optics design, stability of the pointing platform, avoidance of
systematic offsets, and development of data analysis techniques.
Here we describe the experiment, its strengths and weaknesses, and the
various things we have learned that may benefit future efforts to
measure the polarization of the CMB.
\end{abstract}

\section{Instrument Design}
\CO was designed to  measure the polarization of the
CMB on intermediate angular scales.  Here we describe the major
components of the instrument and how it performed in the field.  There
are essentially three major parts: the 
receiver, the optics system, and the pointing platform.  An addendum
concerning the observing site is included as well.  Note that further
details on any part of the experiment discussed in the paper are
available \cite{farese02}.

\subsection{Receiver Description}
The polarimeter implements state-of-the-art HEMT (High-Electron Mobility
Transistor) amplifiers that operate in the 26-36 GHz
frequency range (Ka band).  These amplifiers provide coherent
amplification with a gain of $\approx 25$ dB.  To reject $1/f$ noise
from the detectors and atmospheric fluctuations two HEMTs are
configured as an AC-biased correlation receiver.  Each HEMT amplifies
one of two polarizations observed through the same horn and thus the
same column of atmosphere and same location on the sky at a given
time.  The resultant amplified signals are mixed down to 2-12 GHz,
phase modulated at 1 kHz, multiplexed into three sub-bands, and
further amplified along separate but identical IF amplifier
chains. The signals are then multiplied together and the 1 kHz switch
is demodulated.  The resulting correlated signal time-averaged
amplitude is proportional to one linear combination of the Stokes
parameters (Q or U in the detector's frame) as determined by the
paralactic angle of the observations and the orientation of the
receiver axes.  A second linear combination can be obtained after a
$45 \arcdeg$ rotation of the polarimeter about the horn's optical axis
thus allowing one to measure both Stokes parameters and thus obtain
all information about the linear polarization.  For the observations
reported here no rotation was performed; solely the detector's U
polarization was measured.  Depending on the observation strategy this
will then map to some linear combination of Q and U Stokes parameters,
and similarly ``E'' and ``B'' symmetry modes, on the celestial sphere.
Details for our observations are provided in section
\ref{text:obs_strat}. 

The three sub-bands, in order of decreasing RF frequency, are termed
J1, J2, and J3. Each of these sub-bands is demodulated with waveforms
both in and out of phase with the phase-modulation signal.  Thus the
desired signal is obtained from each in-phase demodulation (called
J1I, J2I, and J3I) in addition to a null-signal noise monitor from the
out-of- phase demodulation (J1O, J2O, and J3O).  Further, a power
splitter prior to the multiplexing stage allows a total power
detection of each linear polarization termed TP0 and TP1.

This receiver was originally coupled to a corrugated feed horn with a
$7 \arcdeg$ FWHM beam for a large angular scale CMB polarization
experiment known as \PO \cite{keating01}.  For further details of the
receiver please see Keating et. al. \cite{keating03}.  In order to
observe smaller angular scales where a larger primordial polarization
signal is expected a dielectric lens was added to the \PO optical
system and this horn+lens combination was coupled to a 2.6-meter
on-axis Cassegrain telescope to form \CO.

\subsection{Telescope Description}
The \CO optics were designed to be as free from systematic effects as
possible.  Oblique reflections of light off metallic surfaces induce
spurious polarization. In the case of a single angle of incidence onto
a flat mirror the induced polarization is given by \cite{renbarger98}
\begin{equation}
P=\big[\frac{\omega \mu}{2 \pi \sigma}\big]^{(1/2)} \sin{\theta} \tan{\theta}.
\end{equation}
Here $\theta$ is the angle of incidence, $\omega$ is the angular
frequency of the incident light, $\mu$ us the magnetic permeability,
and $\sigma$ is the {\bf DC} electric conductivity.
For Aluminum Alloy, 6061-T6, at microwave frequencies these numbers
are typically $\mu \sim 1$ and $\sigma \sim 2.31 \times 10^{17}$
Hz. 

In any off-axis telescope there will be an average angle of incidence
resulting in a systematic polarization for at least one Stokes
parameter.  To avoid this effect \CO implemented an on-axis Cassegrain
configuration. (Note that this systematic polarization can be somewhat
reduced for an off-axis observation  through the use of a fold 
mirror or optics which satisfy the Dragone conditions, though
discussion of these solutions is beyond the scope of this paper.)

\begin{figure}[!t]
  \begin{center}
    \includegraphics{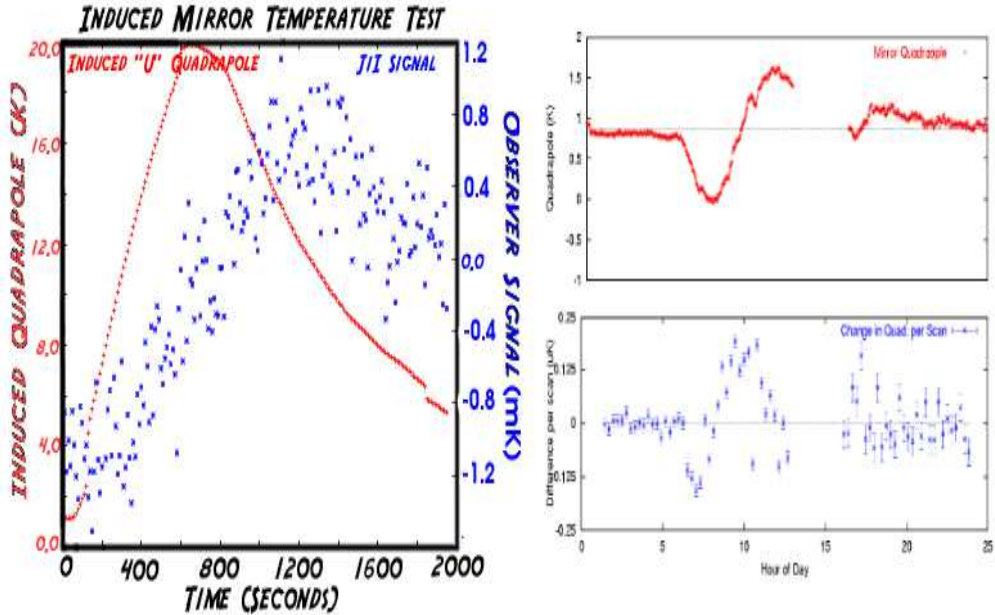}
    \caption[Mirror Quadrupole]{Effects of mirror quadrupolar temperature
      distribution.  The left image shows the calibration data where a
      quadrupole temperature was applied to the mirror with heaters. The top
      right shows the actual quadrupole as a function of time of day,
      as measured by mirror mounted thermometers,on March 25, 2001,
      and the bottom right an estimate of the maximum induced signal
      given the other two plots.} 
    \label{fig:quad}
  \end{center}
\end{figure}

Despite the great reduction in overall systematic polarization in an
on-axis system it is
still possible to obtain a net instrumental polarization for a given
Stokes parameter (expressed in the detector coordinate system).  If
there is a net quadrupolar temperature distribution of the optical
system that is not orthogonal to the observed polarization axes a
systematic polarization will result. Such a primary mirror temperature
profile is likely to arise during periods of
changing thermal conditions such as sun rise and set.  To  understand
how significant this effect could be the temperature 
of the \CO primary mirror was measured at eight locations allowing
decomposition into the dipole and quadrupolar components. We induced
a variety of temperature profiles using mirror-mounted heaters and
measured the resulting polarized offset.  The results of this test are
provided in figure \ref{fig:quad}.  To minimize this effect
during observations  the back of the mirror was insulated with
conventional fiberglass insulation and ``reflectix'' insulation.
An estimate of the maximum induced polarization signal on the
timescale of one scan is given in figure \ref{fig:quad}.  This
allows us to set an upper limit of 0.2 $\mu K$ of induced signal from this
effect during our CMB observations.

Scattering or diffraction by a metal or dielectric will also induce a
polarized signal \cite{kildal88}.  In conventional on-axis systems
metallic or dielectric secondary supports (often termed ``struts'')
necessarily obstruct the optical aperture.  This will give rise not
only to an overall polarization but also an increased side-lobe level.
To avoid both of these effects \CO uses a radio-transparent
expanded polystyrene (EPS) secondary support system.  Finite Element
Analysis (FEA) was used to design this foam support system to minimize
the optical depth while providing sufficient stiffness to position and
stabilize the secondary mirror with 1 mm resolution.

Many types of foam were measured for their emissivity and reflectivity
at Q (36-44GHz) and W (88-102 GHZ) bands.  Our final candidates were
various densities of EPS. A conical shape was
chosen for this support structure with the additional of metallic
rings at the center and periphery of the cone to add
stiffness. Neither of these rings is visible to the detector. As a
trade off between maximizing the stiffness of the structure and
minimizing its optical depth EPS of two pounds per cubic foot was
selected.  

\begin{figure}[!t]
  \begin{center}
    \includegraphics{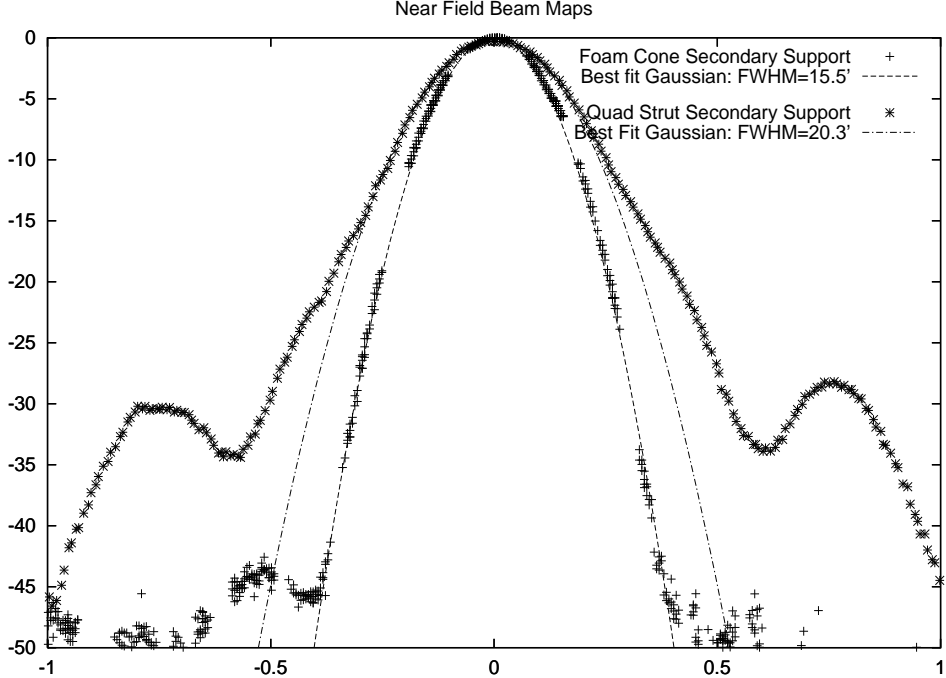}
    \caption[Strut Effects]{Beam maps of the \CO main lobe and first
      sidelobes using a mid-field source in two different secondary
      support configurations: EPS  and struts.  Note that the larger FWHM
      of the strut configuration is because the source was closer to the
      telescope resulting in an effectively larger beam.  Note also that the
      first sidelobe is reduced from -30dB to -45 dB through use of the
      EPS support.} 
    \label{fig:side_lobes}
  \end{center}
\end{figure}

The success of this approach was tested through beam maps of the
system using a Gunn oscillator located in the mid-field for each of two
configurations: support of the secondary mirror with conventional,
optimized metallic struts and with the foam support.  Use of this cone
when compared to the conventional optimized strut configuration
reduced the first sidelobe from -30 dB to -45 dB as shown in figure
\ref{fig:side_lobes}.

\begin{figure}[!t]
  \begin{center}
    \includegraphics{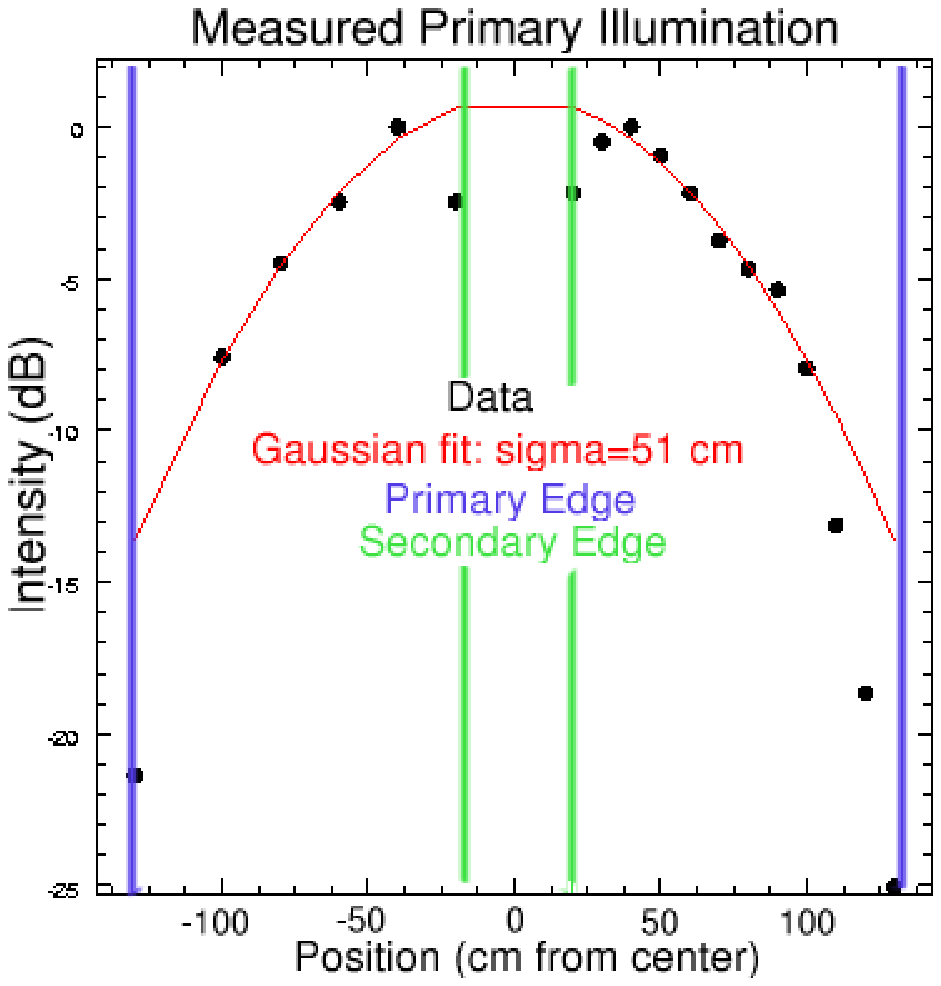}
    \caption[Primary Illumination]{A map of the measured primary
      mirror illumination vs. distance from the center.The rapid edge
      taper and effect of the central hole are readily apparent at
      large and small radii, respectively.}
    \label{fig:prim_ilum}
  \end{center}
\end{figure}

The secondary mirror was designed to minimize aperture blockage.  The
primary mirror that was acquired had a hole of 30 cm diameter in its
center, thus the secondary was constructed with a 30 cm diameter.  A
hole was left in the center of this mirror to prevent re-illumination
of the receiver and a polarized IR calibrator placed behind this hole.
To reduce edge illumination of the primary the edge of the secondary
mirror is designed to illuminate a surface 7 cm in from the edge of the
primary thus giving a much faster than Gaussian taper to the edge
illumination.  This results in reducing ground pickup and spillover
while still maintaining a small beam size.  
The illumination of the primary was measured to be less than -25dB at
the edge.  A one dimensional plot of the primary illumination pattern
is provided in figure \ref{fig:prim_ilum}.  

In order to minimize the illumination at the edge of the secondary
mirror using the existing microwave horn and dewar it was
necessary to reduce the beam size from 7 degrees FWHM to 5 degrees
FWHM with a lens. The
additional requirement that this lens be cooled to reduce its
contribution to the system temperature necessitated that the lens be
mounted close to the horn and thus be approximately the same diameter as
the horn.  A simple meniscus phase-correcting lens was selected which
resulted in -15dB secondary edge illumination.  This design was based
on the work by \cite{kildal88} and has a spherical inner surface,
whose radius matches that of the radius of the horn-produced Gaussian
beam, and an ellipsoidal outer surface designed to give a flat phase
front at the entrance surface of the lens.

\begin{figure}[!t]
  \begin{center}
    \includegraphics{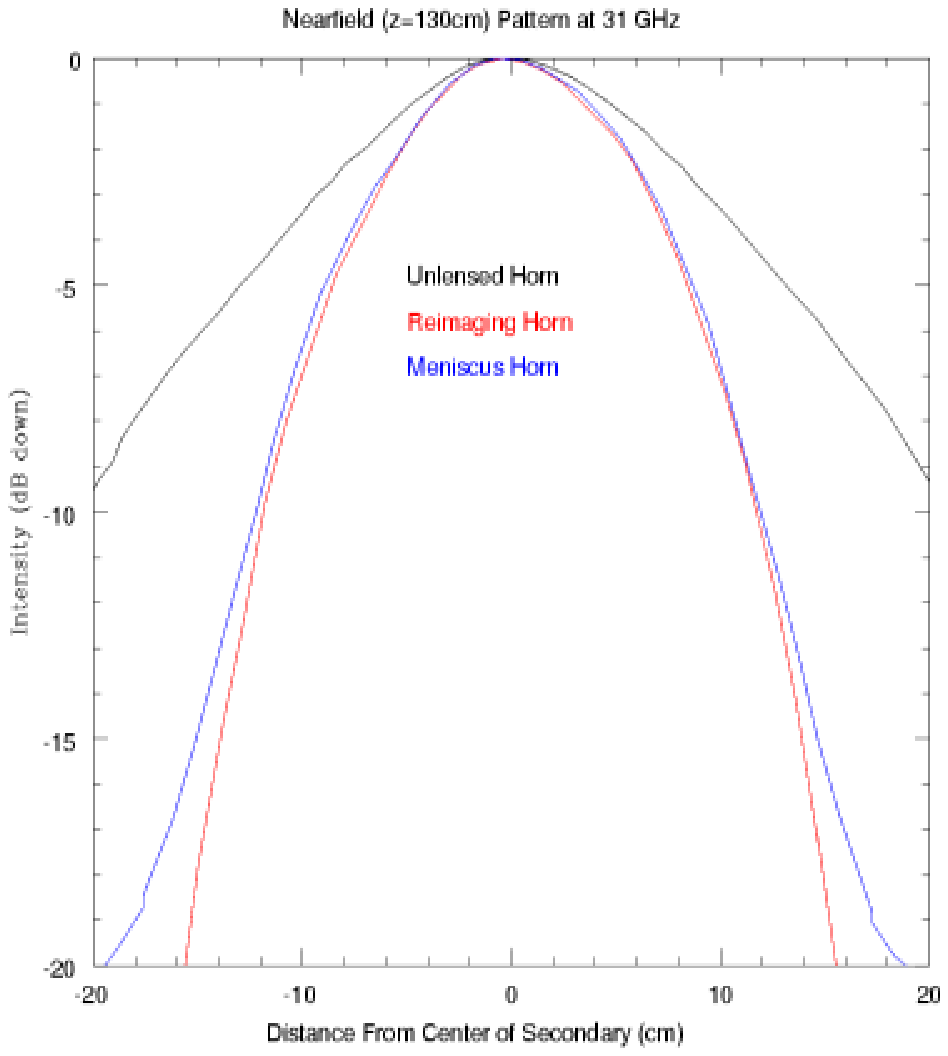}
    \caption[Lens Comparison]{Comparison of the beam patterns of meniscus
      and bi-convex lenses at the distance of the \CO secondary mirror.}
    \label{fig:lens_comp}
  \end{center}
\end{figure}

A second design that was considered was a biconvex lens.  Using
Gaussian optics theory \cite{goldsmith98} we
can consider this lens as reimaging the beam waist of the horn.  The
radii were chosen to reimage the waist a distance d in front of the
horn with d also the lens to beam waist distance.  This provides the
same far field resultant beam but would give a lower edge illumination
at the secondary as shown by figure \ref{fig:lens_comp}.  Because the
theory, focal point, and beam shape were better understood for the
meniscus lens, however, it was selected for use.  Though this resulted
in a more aggressive illumination of the secondary than desired most
of the excess power comes from the sky and thus was not anticipated to
produce significant systematic effects.  Unfortunately, systematic
effects were observed and this aggressive illumination may well be the
cause.

To further reduce \CO' sensitivity to possible systematic effects two
levels of ground screens, one affixed to the telescope and the other
stationary, were constructed.  The screens which were mounted to the
telescope were attached directly to the edge of the primary mirror.
These provide an additional $> 50$ dB of attenuation to signals from
the ground, sun, and moon in addition to the already low sidelobe
level (-60 dB) of the telescope at the location of these contaminants.
Observations were conducted both with and without the outer
(stationary) ground screens present.  Data collected with the outer
groundscreens present suffered from a larger scan synchronous signal (SSS)
than data taken with them absent.  It is believed that the combination
of telescope spillover with the oblique scattering angle off the
metallic surface of the groundscreens induced a large polarized offset.
This will be discussed in more detail in section \ref{text:SSS}.
{\bf This demonstrates that even more than exceptional care must
  be given to designing a CMB polarization experiment to be immune to
  systematic  effects.} 

\subsection{Platform Description}
\CO is based on a standard Az-El pointing platform.  The azimuth
and elevation stages are separate units.  The azimuth table used was a
refurbished and improved version of an existing table previously
designed for use with CMB observations from the South Pole.  The
elevation stage was designed and built for this experiment. 

The table consisted of a 3/8'' thick Aluminum plate reinforced with 1.75''
long fins protruding from the bottom.  These fins attached to a
conically shaped plate which rode upon four conical roller bearings
which supported the weight of the telescope.
A 12'' diameter roller bearing at the center connected
the table to the base.  The base was composed of light-weighted
Aluminum beam and a 1/2'' thick plate.  Wheels and screw jacks were
attached to this base allowing either mobility or stability.  The
table was, however, designed for a 1.3 m dish and its associated load
and thus proved to be insufficiently stiff to support the \CO
mirror. Several methods of improving the  table were investigated. 
The simplest satisfactory result involved rigidly bolting the elevation 
base's 3'' Aluminum box beam to the table. After this modification the
table deflected less than 1/16'' and did not tilt (i.e. $< 1^{'}$)
when scanning in Azimuth as per our scan strategy (see section
\ref{text:obs_strat} for details).  Further, the rotation platform's 
support structure was replaced with a 12'' Aluminum I-beam (W12x7)
triangle with a 1/2'' thick Aluminum plate welded to its top.  The
angular deflection under 1000 ft-lbs for this base is much less than
$1'$.

The table was actuated by a stepper motor driving a harmonic drive
gear reducer which rotates one of the 4 conical bearings described
above.  The harmonic drive has a gear reduction of 60:1 and the ratio
of the radius of the bearing to that of the surface it drives yields a
further reduction of 6:1 resulting in a 360:1 net gear reduction.
Using our 200 step/revolution, 5 ft-lb motor allows us a single step
resolution of better than 20 arc seconds and the ability to drive
torques up to 1800 ft-lbs.  Unfortunately the effective torque
specification of this stage was limited by the friction between the
bearings and the conical plate.  In periods of high wind pointing
control was lost and the associated data ignored.  Hard stops were
implemented which prevented motion of more than a few degree outside
the desired scan region.  One important lesson to learn from this,
however, is that {\bf friction drive based systems can not apply
sufficient torque to control a mid-size telescope in periods of even
moderate wind and should be avoided.}

The mechanical coupling of this drive system was improved by
firmly bolting all members and removing backlash from the harmonic
drive with shim material.  The compliance and backlash were tested by
projecting a 1/8'' diameter laser beam mounted to the base to a wall
100 feet away.  A backlash of less than 1.1' and a compliance of 1.5'
were observed using this method which was later confirmed when
performing beam maps to test the optical system.  These specifications
are sufficient for our 20' beam.

The elevation axis was driven by a ball-screw linear actuator  
of 60'' length when extended with 26'' of stroke to allow
observations at any elevation.  The actuator was driven by a 40:1
worm gear which was coupled to a reducer and stepper motor.  The pitch
of the worm gear assures that the actuator cannot be driven backward,
the ball-screw grants low backlash and compliance, and the stepper
motor and reducer are chosen to provide sufficient resolution and
speed. 

The position of each axis was read out by a Gurly 16 bit encoder giving a
resolution of 20 arc seconds. The mounting locations of the encoders were
chosen so that they supported no torque thus assuring accurate readout
of the coordinates. Data acquisition of all radiometer, pointing, and
``house keeping'' data (such as thermometry) as well as telescope
control were performed with a Pentium class laptop computer and 48
channel 16-bit ADC purchased from Iotech.

Our campaign took place at the University of Wisconsin's Astronomy
observatory in Pine Bluff, Wisconsin (89.685 West longitude, 43.078
north latitude) 100 meters from the platform used for the \PO
observations.  As a ground-based experiment located within twenty
miles of the UW main campus \CO had the advantage of being deployed
and staffed relatively easily.  Though some sensitivity and observing
time is sacrificed because of poor atmospheric conditions, the
advantage of being able to repair and adjust the telescope during the
course of the observations was invaluable.  The site chosen used a
60'x30' tensioned fabric building with a wheeled, aluminum frame to
house the telescope \cite{BigTop}.  This building was rolled 20 m to
the South of 
the telescope on tracks for observations and rolled over the telescope
for suitable shelter during periods of foul weather.  By moving the
building rather than the telescope we were assured of the stability of
the telescope and its celestial alignment.

\section{Instrument Characterization}
Once deployment began in mid-January of 2001 a number of crucial
tests needed to be performed to understand the performance, and
results, of  the experiment.  These include the calibration, beam
characterization, and pointing of the telescope.  We also include here
a discussion of our scan strategy.

\subsection{Calibration}
There are three bright objects visible in the northern hemisphere with
sizable average polarization and angular size of less than several
arc-minutes.  These are Tau A (The Crab Nebula), Cas A, and Cyg A with
polarized signals at 30 GHz of $\sim 23.5$ \cite{johnston69}, 1.94
\cite{mason99}, and 
1.1 Jy \cite{melhuish99}, respectively.  With our $19.6^{'}$ beam and the
noise in our correlator channels, these correspond to a signal to
noise ratio in one second of observation of $\sim$ 20, 2, and 1.
Clearly Tau A is our primary source candidate for calibration.

\begin{table}[!t]
  \begin{center}
    \begin{tabular}{|c|c|c|c|} \hline
      Factors	   &  J3I  & J2I  & J1I            \\ \hline
      Ant. Temp. (mK) & $11.8 \pm 2.5$&
      $9.6 \pm 2.1 $ & $7.7 \pm 1.7$          \\ \hline
      inv. Tele. eff  & \multicolumn{3}{|c|}{2.68} \\ \hline
      atm.	   & \multicolumn{3}{|c|}{0.988}   \\ \hline 
      RJ, CMB	   & 1.020 & 1.024 & 1.029         \\ \hline
      cosmo/astro& \multicolumn{3}{|c|}{0.5}  	   \\ \hline
      decay 	   & \multicolumn{3}{|c|}{0.947}   \\ \hline
      Signal (mV)& $12.4 \pm 0.37$ &
      $5.15 \pm 0.14$ & $3.59 \pm 0.16$            \\ \hline
      Gain (K/V) &  $1.2 \pm 0.4$ &
      $2.4 \pm 0.9$ & $2.8 \pm 1.2$            \\ \hline
    \end{tabular}
    \caption[Calibration]{Provided are the various calibrations for our 3
      correlator channels and 
      two total power channels derived from observations of the Crab
      Nebula. ``atm.'' indicates corrections for atmospheric absorption,
      ``RJ, CMB'' for the conversion from Raleigh-Jeans to thermodynamic
      temperature, ``cosmo/astro'' is described in the text, and ``decay''
      for the decay of source brightness since the measurements of
      \cite{johnston69}.} 
    \label{tab:cal}
  \end{center}
\end{table}

The present day flux for Tau A and our expected signal are calculated
using the polarized observations of Johnston and Hobbs
\cite{johnston69} adjusted for the observed decay rate of Allers
\cite{aller85}, and spectral index of Baar \cite{baar77}.  Johnston
and Hobbs measured a flux of $313 \pm 50$ Jy at a frequency
of 31.4 GHz with 8.1\% polarization at a polarization angle of
$158^o$. The flux of Tau A is known to be decreasing with time at a rate of
$0.167\% \pm 0.015 \%/$year \cite{aller85}.  The flux density spectral
index is measured to be $\alpha=-0.299 \pm 0.009$ \cite{baar77} so the
polarized antenna temperature is taken to be
$P_{Jx} \propto (\nu_{GHz}/31.4)^{-2.299}$.   Several
corrections are performed as summarized in table
\ref{tab:cal}.  Most notable is a change in convention between
cosmologists and astronomers and  is obtained by dividing by
two \cite{odell03}. This later correction is driven by the Rayleigh-Jeans
law where we take:
\begin{equation}
S=2 \frac{k_B T}{\lambda^2} 
\end{equation}
The factor of 2 arises because 
there is radiation in both polarization states.  Thus if one is
calculating the total flux of a source it is consistent to define:
\begin{equation}
S=\frac{k_B}{\lambda^2} (T_x+T_y)
\label{eq:S}
\end{equation}
where $x$ and $y$ are the two measured polarization states.  From a
radio astronomer's point of view one would report Tx+Ty allowing
direct conversion to flux units.  In order for the CMB cosmologist to
maintain consistency with equation \ref{eq:S} therefore would dictate
that one define $2 T=T_x+T_y$, or similarly:
\begin{eqnarray}
T_I = \frac{(T_x+T_y)}{2}, \\
\text{and therefore also} \nonumber \\
T_Q =\frac{(T_x-T_y)}{2}.
\end{eqnarray}
We have added the subscript $I$ and $Q$ to denote the total intensity
and ``Q'' polarization temperatures respectively.  Our definition is
in agreement with existing CMB polarization experiments and is
consistent with CMBFAST \cite{uros_pc} where one requires that $T_I$
and $T_Q$ have the same numerical prefactor.  The expected signal and
all necessary corrections as a function of channel is shown in table
\ref{tab:cal}.  We are currently finalizing the calibration by
considering Tau A observations by other observers.

\subsection{Pointing}
We initially pointed the telescope by co-aligning an optical telescope
mounted on the primary mirror with the cm-wave beam pattern using a 31
GHz Gunn oscillator mounted on a radio tower 1.9 km West-by-Southwest
from the telescope (see section \ref{text:tower} for details).  The
source on the radio tower was  quite easily
visible with the optic selected.  Because of the proximity of our
observing region to the NCP several optical observations of Polaris
and a number of nearby stars were made.  These Polaris observations
were then used to define our absolute azimuth and elevation offset and
thus to define our observing region.  Unfortunately, misalignment of
the optical telescope resulted in an azimuth and elevation pointing
error of $+0.458 \pm 0.027$ and $-0.107 \pm 0.023$ degrees
respectively.

This pointing offset was discovered through observations of the
supernova remnant Cas A.  Cas A is circumpolar from our observing
location, fairly close to our 
observing region, and quite bright in intensity, making it useful as a
pointing tool.  Three observations were made throughout the season as
outlined in table \ref{tab:casA}.  Through linear fits the elevation
is seen to drift at $-1.43 \pm 2.3$ arcminutes in Elevation and $10.1
\pm 6.9$ arc minutes in Azimuth per month; There is no compelling
statistically significant evidence for drifts in the pointing over the
given 52 day period.  Thus, a fixed pointing offset was used for all
files in the analysis.  {\bf This also gives us confidence that EPS
secondary supports are indeed stable enough, and transparent enough,
to be used in precision observations.}  

\begin{table}[!t]
\begin{center}
\begin{tabular}{|c|c|c|c|c|c|} \hline
Day & Obs. & Az   & error &  El   & error \\ \hline
62  & 4    & 17.4 & 6.6   & -5.4  & 4.2 \\ \hline
109 & 2    & 40.8 & 1.8   & -10.2 & 2.4 \\ \hline
114 & 6    & 29.4 & 4.3   & -6.0  & 1.8 \\ \hline \hline
ave &      & 37.8 & 1.6   & -7.3  & 1.4 \\ \hline
RMS &      & 12.6 &       &  2.1  & \\ \hline
wRMS&      & 5.3  &       &  1.8  & \\ \hline
\end{tabular}
\caption[Cassiopeia A: Pointing]{Successful observations of
Cas A were made on the days specified by the day number in the
first column.  Obs. gives the number of full, independent rasters
performed at that time.  The azimuth and elevation offsets and errors
are given in the remaining columns.  ave provides the weighted
average, and thus the offsets used in analysis, RMS the scatter of the
individual measurements, and wRMS the scatter weighted by the
uncertainty of each measurement.}
\label{tab:casA}
\end{center}
\end{table}

\subsection{Beam Determination}
\label{text:tower}
Raster maps of the above-mentioned ``tower source'' were made by
maintaining constant elevation while scanning in azimuth with
elevation changes performed at turn-around points.  This ``tower
source'' was a 31-GHz Gunn oscillator mounted on a radio tower 1.9 km
West-by-Southwest of the telescope.  This was essentially in the far
field of the telescope requiring only a 0.1' correction to the
beamwidth at infinity.  The source was mounted on the tower  and could
be turned on and off with a very long extension 
cord; thus allowing it to be bright enough for our use as a tool but
not worrisome as a signal contaminant during observations. This Gunn
oscillator was mounted to a scalar feed horn in a 
sealed PVC tube with foam endcaps.  Thus it was sealed against weather
and quite easily visible with the optical telescope which was mounted
on the primary mirror and thus coaligned with the optical beam.

The FWHM obtained by fits of a Gaussian to the elevation and azimuth
directions for two separate days of tower source observations are 19.2
$\pm$ 0.4 and 20.3 $\pm$ 0.5 arcminutes respectively.  These beam maps
are confirmed by scans of Tau A.  A fit of these maps with Tau A
deconvolved yield a FWHM for each polarization channel and each
direction ( elevation and cross-elevation) as provided in table
\ref{tab:FWHM_Tau}.Note that the analysis of the Tau A
data requires time series  filtering which may induce a larger beam
size. These numbers are further confirmed by observations of Venus
which is a point source in our beam and yields beams of the two total
power channels of $18.5^{'} \pm 1.0^{'}$  and $19.6^{'} \pm 0.8^{'}$.
In our likelihood analysis we make the approximation that the beam is
axially symmetric with FWHM $20.0^{'}$ an approximation which affects our
result by $\le$ 5\%.

\begin{table}
  \begin{center}
    \begin{tabular}{|c|c|c|} \hline
      Channel     & Cross-Az & elevation \\ \hline
      J1I (32-35) & $21.5 \pm 0.9$ & $20.2 \pm 0.9$ \\ \hline
      J2I (29-32) & $22.0 \pm 0.7$ & $21.5 \pm 0.9$ \\ \hline
      J3I (25-29) & $23.2 \pm 0.7$ & $23.2 \pm 0.9$ \\ \hline
    \end{tabular}
  \end{center}
  \caption[Beam Widths]{ FWHM of the beam on the sky of each sub-band 
    derived from observations of Tau A, in arcminutes.  All are biased
    somewhat high because of the data processing performed}  
  \label{tab:FWHM_Tau}
\end{table}

\subsection{Observing Strategy} 
\label{text:obs_strat}
Our observing strategy was designed to optimize the probability
of detecting a signal under current well-motivated theories while
still allowing for systematic checks and tests.  We attempted to
observe a circular ``disk'' region centered on the North Celestial Pole (NCP)
by maintaining a constant elevation and scanning the telescope in
azimuth.  During constant-elevation scans the thermal load from the
atmosphere remains constant and reduces both gain fluctuations of
the receiver and  intensity-polarization coupling in the polarimeter
that could cause systematic effects.  Our full scan period 
was varied between 10 and 20 seconds to reject 
longer term atmospheric fluctuations while maintaining stable and
reliable telescope performance.  

\begin{figure}[!t]
\begin{center}
\includegraphics{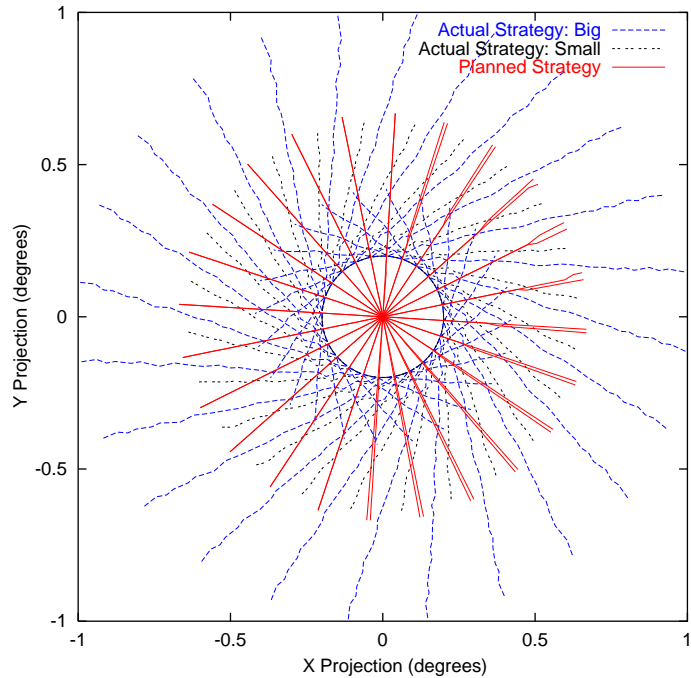}
\caption[Scan Strategy]{\CO scan strategy in rectangular projection
  about the NCP.  The axis are real degrees on the sky with the positive x-axis
  corresponding to a Right Ascension of 0.  It is clear
  how constant Elevation Azimuth scans are transformed into a two
  dimensional map through sky rotation.} 
\label{fig:ScanStrat}
\end{center}
\end{figure}

As the sky rotates this scanned line is transformed into a cap
centered on the NCP as demonstrated in figure \ref{fig:ScanStrat}.
Further, because of the sky rotation, each half
sidereal day the same region of sky is observed and allows the use of
difference maps as a robust test of systematic errors.  Initially this
``cap'' was one degree in diameter in order to allow deep
integration on a small patch of sky to search for systematic effects.
Half way through the season this diameter was increased to 1.8 degrees
to allow a trade-off between sample variance and noise-induced
errors.  As mentioned above there was a small pointing offset so our 
actual scan strategy was not the one we intended.  This resulted in a
loss of symmetry and thus some systematic tests but acceptable noise
properties.  Given the relative sizes of our beam, scan region, and 
pointing offset from the NCP our telescope is sensitive to both ``E''
and ``B'' symmetry modes at roughly the same
level. \label{text:scan_strat}

One further benefit of this scan strategy is that it simplifies the
process of encoding the time stream data into a map which displays
well the SSS while still allowing for reasonable cross-linking, sampling
of a variety of paralactic angles and excellent systematic tests.
As mentioned below,
a one dimensional to two dimensional map comparison allows simple
identification of SSS.  If we plot out data in a coordinate system
defined as Azimuth and Right Ascension it becomes quite easy to
combine the data across many days and project out the modes that
correspond to SSS (see section \ref{text:SSS_modes}).  

Finally two configurations of the ground screens were implemented; one
utilizing the two layers mentioned above and one with only the inner,
co-moving ground screen.  These two options, combined with the two
scan amplitudes mentioned above, result in a total of four sub-seasons of
data termed SIGS, SOGS, BIGS and BOGS to specify whether the (S)mall
or (B)ig scan was used and weather the (O)uter or only (I)nner ground
screens were in place.  Analysis was performed on each channel in
each sub-season as well as combinations of channels within a
sub-season, combinations of sub-seasons for a given channel, and all
acceptable data. 

\section{Data Characterization}
Here we give a preliminary glance at the data emphasizing our
selection process, data characterization, statistical checks, and
analysis pipeline. 

\subsection{Data Selection}
There were a total of 1776 hours available during our observing season
which was defined as the months of March, April, and half of
May of 2001. Of these 409 were sufficiently good weather to operate the
experiment.  74 hours were ignored because heavy winds disrupted
the Azimuth pointing and control and 28 hours were removed because of
equipment failures.  This leaves a total of 309 hours of usable data
observing the target region.  Further cuts are made to select periods
of stable observing conditions.

The data are divided into files of 15 minutes in length.  For each
file six statistics are generated: three 
noise-based statistics (white noise level, 1/f knee, and $\zeta$
\cite{keating01}); two cross correlations (between two correlator channels
and between a correlator and total power channel); and a linear drift of
the time stream.  The 1/f knee and $\zeta$ statistics proved to be
most sensitive to periods of light cloudiness or haziness.  The cross
correlations were most useful in identifying more rapid contaminating
events such as discrete clouds, birds, or planes interfering with the
observations.  Finally the linear drift was used to identify dew
formation on the foam cone.

This dew formation was the one drawback of the foam secondary
support.  This problem is conceivably circumventable with the use of IR
lamps, fans, or some similar technology.  Fortunately it was quite
easy to identify periods of excessive dew formation as the correlator
outputs would drift up as the dew formed and down as it evaporated.
Once this data was cut using the linear fit to the time stream there
is no evidence that such periods further corrupt the data.

\begin{figure}[!t]
\begin{center}
\includegraphics{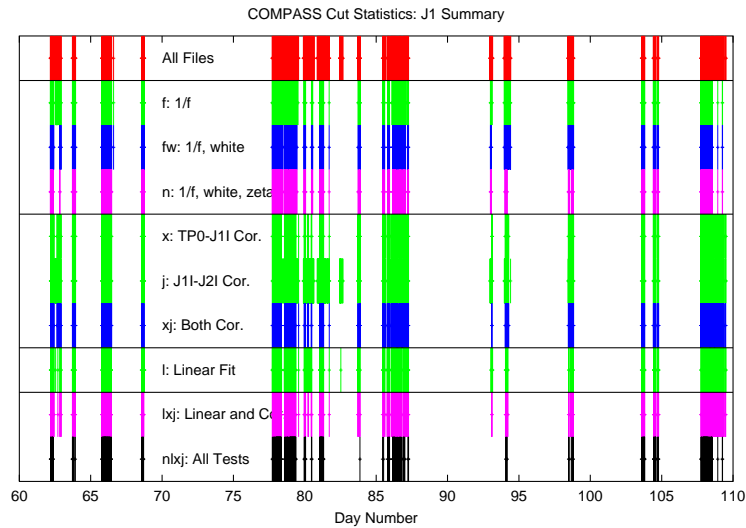}
\caption[Data Cut Summary]{\CO data selection summary for one channel
  (J1).  Time proceeds along the x axis with each vertical line
  representing a 15 minute file that passed  the indicated cut.  Green
  lines indicate that only a single cut was performed while blue indicate
  multiple cuts were used. Magenta and black indicate cuts that were
  used in actual full analysis runs.}
\label{fig:cuts}
\end{center}
\end{figure}

A histogram of each statistic is formed and 
unions and intersections are made of files passing sets of cuts at the 
3-$\sigma$ level.   Our results were not overly sensitive to which
sets were selected.  The results reported here use the two
cross-correlation criteria for they retained the most data while still
passing all null-tests performed (see below).  A pictorial summary of the
cuts considered is shown in figure \ref{fig:cuts}

Each file is also passed through a despiking procedure.  Regions with
excessive slopes, second derivatives, or values greater than 5 standard
deviations from the mean are flagged, cut, and filled with white noise as
estimated by the remainder of that file.  This procedure removes
between 0.1 \% and 32\% of the data from any give sub-season. The 32\%
was an anomaly; it occurred only for one channel,
J3, during the first sub-season where a damaged preamplifier gave rise
to excessive spikes in the data intermittently.  This was fixed once
the cause was identified.  After all selection procedures the number
of hours kept were 144, 123, and 164 for J1, J2, and J3 respectively.

\subsection{Scan Synchronous Signal}
\label{text:SSS}
One common systematic error in scanning style experiments is the
presence of non-celestial signal that correlates highly with the scan,
termed here ``Scan Synchronous Signal'' (SSS).  In \CO such SSS was
observed and was related to a polarized offset induced by oblique
reflection from the stationary ground screen or spillover to the
ground.  In all sub-seasons a variable offset and linear term (when
plotted against Azimuth) are observed and removed.  In the larger scan
with the stationary ground screen present a quadratic signal is
detected and removed as well.  Once these removals have been performed
the residual is consistent with Gaussian noise.

Detection of the SSS is performed easily by comparing maps made by
binning the data in the (one) scan dimension (i.e. Azimuth for \CO) to
those made by binning in two (i.e. Azimuth and Right Ascension).  If
one has SSS contamination the relationship between the reduced
chi-squared of the one dimensional map to that of the two dimensional
map is given by
\begin{equation}
\chi^2_{1D}=1+N_{\text{bin}} \times (\chi^2_{2D}-1),
\end{equation}
where $N_{\text{bin}}$ is the number of bins in the dimension that is not
present in the one dimensional map, and $\chi$ refers to reduced chi
squared for each map.  As shown for a typical channel in table
\ref{tab:1D2D} there is no 
evidence for SSS remaining in any of the sub seasons after removal of
a first order polynomial other than the BOGS configuration.

\begin{table}
\begin{center}
\begin{tabular}{|c|c|c|c|c|c|} \hline
& \multicolumn{2}{|c|}{Two-D maps} 
& \multicolumn{2}{|c|}{One-D maps} \\ \hline
Sub-Season&DOF's& $\chi^2$&DOF's& $\chi^2$ \\ \hline 
SIGS & 244 & 1.00 & 13 & 0.76 \\ \hline
SOGS & 232 & 1.09 & 13 & 0.45 \\ \hline
BIGS & 182 & 1.08 & 18 & 0.93 \\ \hline
BOGS & 217 & 1.24 & 19 & 3.96 \\ \hline
\end{tabular}
\caption[Map $\chi^2$ Tests]{The Chi-squared tests for RA-Az and RA-Az
binned maps for all sub-seasons for channel J1I.  Note that all are
consistent with no signal except BOGS which does show
clear signs of unremoved Scan Synchronous Signal.}
\label{tab:1D2D}
\end{center}
\end{table}

In order to be insensitive to this contaminant we remove a polynomial
fit (second order for BOGS data, first for all others) from each
Azimuth scan (roughly ten seconds of data) for each channel. When
producing the noise covariance and associated likelihood we then
must, and do, apply the constraint that this mode be ignored (see
section \ref{text:SSS_modes}).

\subsection{Map Making}
In order to estimate a power spectrum from our data it is necessary to
produce a map of our results.  By the term ``map'' here we mean a
pixelized representation of the data which contains information on the
spatial location, most likely data value and (non-diagonal) noise
correlation matrix.  The nature of polarized observations
requires different mapping procedures than a simple intensity map.
Traditionally either a map of polarized intensity and orientation or a
map of the Q and U Stokes parameters are provided.  As \CO observed at fixed
elevation the polarization direction information is unambiguously
encoded in the Right Ascension and Azimuth. 

{\bf In investigating how to produce a map of this data we discovered
that using a variety of maps is a very useful tool for data
visualization and discovery and removal of systematic effects.}  We
map initially each data file in Azimuth coordinates as the sky
rotation on this time scale is sufficiently small that a single bin of
RA is needed.  This intermediate product is quite elegant in that 
it allows large data compression without information loss, facilitated
the process of encoding the constraint matrices, and proves flexible
for producing a variety of other maps for systematic tests.

\begin{figure}[!t]
\begin{center}
\includegraphics{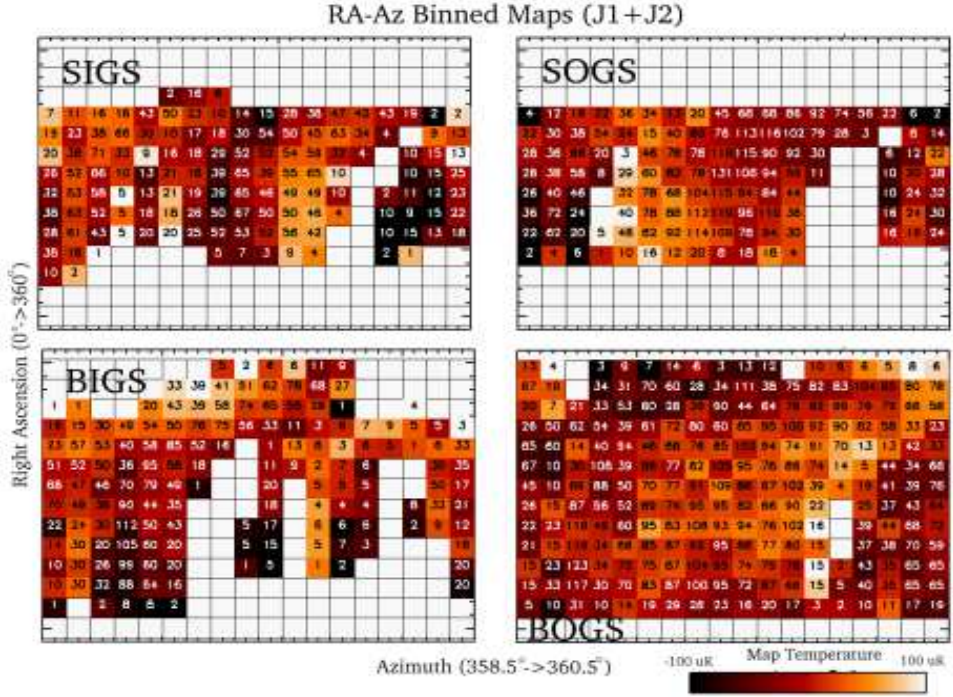}
\caption[RA-Az Maps]{Maps plotted in Right Ascension and azimuth
coordinates.  Shown are maps for all 4 sub-seasons of the data for J1I
and J2I coadded. Color indicates intensity
of the signal per pixel while the number written over that pixel is
the number of Az binned files that contribute to the value.}
\label{fig:maps}
\end{center}
\end{figure}

One useful map format in which to display the data is RA-Az
coordinates. This facilitates identification of systematic effects and
spurious (or real) signals. A second map format is a three dimensional
map of RA, dec, and paralactic angle.  This second map, though
sparsely populated, retains the polarization information of our
observations and is used in power spectrum estimation, though it is
less intuitive to display and interpret.  Maps of the
first style  are provided in figure \ref{fig:maps}.  

\section{Data Analysis}
As mentioned above, each file has an Az ``map'' composed of a data
vector, $\vec{d_i}$, noise covariance matrix ($C_{n,i}$), ``constraint
matrix'' ($C_{c,i}$), and pointing matrix as per the formalism of
Bond, Jaffe, and Knox \cite{bond98}.  The noise
covariance matrix is estimated from the timestream data.  The off
diagonal elements of this matrix were shown to be much less than
$10^{-6}$ the amplitude of the on diagonal elements and are therefore
ignored.  This is because the scan rate is much slower than the
anti-aliasing filter knee.  This indicates that we are not making
optimal use of our frequency bandwidth, but given the stability of the
correlation receiver such optimization was unnecessary.

\label{text:SSS_modes}
The constraint matrix encodes the SSS removal
process and projects the removed modes out of the map by adding noise
of amplitude $10^8$ that of the noisiest on-diagonal element (units of
$\mu K^2$) to the contaminated modes.  These contaminated modes are
constructed for each file's Az-map and for each order of the fit
removed by forming a constraint vector and taking its outer
product. As an example, consider a three element map from which we
wish to remove a constant and a slope.  The constraint vectors are:
\begin{equation}
\begin{pmatrix}
1 \\ 1\\ 1\\
\end{pmatrix}
\text{and}
\begin{pmatrix}
1 \\ 0\\ -1\\
\end{pmatrix}
\end{equation}
with the corresponding constraint matrices:
\begin{equation}
\begin{pmatrix}
1 & 1 & 1\\ 
1 & 1 & 1\\ 
1 & 1 & 1\\ 
\end{pmatrix}
\text{and}
\begin{pmatrix}
1 & 0 & -1\\
0 & 0 & 0 \\
-1 & 0 & 1\\ 
\end{pmatrix}
\end{equation}

These constraint templates are then multiplied by the indicated
(large!) noise and added to the noise covariance matrix of each file
Az map.  These $i$ sub-maps are then added together in
RA-Az-parallactic angle space and a resultant generalized covariance
matrix, $C_N$, and map of the data, $D$, formed as
\begin{eqnarray}
C_{N,i}  &=& C_{n,i}+C_{c,i}, \nonumber \\
C_N^{-1} &=& \sum_i^M C_{N,i}^{-1}, \\
\vec{D} &=& C_N \left[ \sum_i^M \vec{d_i} C_{N,i}^{-1}. \right]
\nonumber
\end{eqnarray}
The  RA-dec-PA angle map was chosen for it eliminates pixelization
effects for our beamsize and provides the most uniform pixel noise of
the methods considered.

As a side note we studied the effect of removing these polynomials in
this way from the data.  Subtraction of a constant term resulted in
weakening of our result (i.e. increasing the uncertainty of the
likelihood or 2-$\sigma$ limit) by 25\%.  Removing a DC level and a slope
weakened it by 50\%!  In retrospect this could have been simulated
prior to the decision on our scan strategy.  Our scan strategy was
designed to give nearly equal noise per pixel on a number of pixels
that minimized the error resulting from the combination of sample
variance and statistical errors.  {\bf It would have been more
powerful to optimize our observing strategy to obtain the greatest
number of signal-to-noise eigenmodes with eigenvalues $\approx$ 1.}
This analysis could have been executed including the effects of
possible systematic effects.

Once these maps were obtained we proceeded to calculate a
likelihood.  A flat band power model of E-mode polarization appodized
by our $20.0^{'}$ beam was used to generate the theory covariance matrix,
$C_T(E)$, though use of a concordance model polarization spectrum did
not significantly change our results.  The likelihood of the amplitude
of this flat band power is calculated using the signal to noise
eigenmode method as described in \cite{bond98}.  To reduce the time to
compute the likelihood curve only modes with an eigenvaue more than
$10^{-8}$ that of the most well measured mode are retained.  It is
worth noting that no eigenmode has an eigenvalue greater than 1.  We
build the total covariance matrix $C(E)=C_T(E)+C_N$ and compute the
likelihood as given by
\begin{equation}
L(\vec{D} | C(E))=\frac{1}{(2 \pi)^{N/2}} 
        \frac{e^{-\frac{1}{2}\vec{d}^T C(E)^{-1} \vec{d}}}
        {\left| C(E) \right|^{1/2}}.
\end{equation}
This likelihood is calculated on a grid spaced uniformly in units of
$\mu K^2$ to avoid over weighting large values of the fit parameter
and is allowed to take values of negative power to test the
correctness of the noise covariance matrix.  {\bf Note that in a low
signal to noise experiment these steps, as well as computing the
actual likelihood curve rather than an estimator, are crucial.}  When
excessively negative values of power are considered one is considering
unphysical solutions and the covariance matrix will no longer be
positive definite.  If the matrix becomes non-positive definite at
some (negative) value of power that value is given a Likelihood of
zero.

Software used to estimate the power spectrum was extensively tested
using dozens of simulated maps generated from a known power spectrum by an
independent piece of software.  Each map was given hundreds of noise
realizations and the statistical properties of the results studied
extensively.  The time stream analysis, map making,
and likelihood codes were all tested independently and as a unified
pipeline.  Maps with no signal as well as those with a signal of a
known spectrum and a variety of amplitudes were considered. After much
effort the software passed all these tests.  It should be noted that
in this testing process we were reminded that, especially in the low
signal to noise range, a likelihood estimator is a biased
estimator.  Our tests required a calculation of the full likelihood
curve rather than the peak and curvature to obtain the correct
statistical results.

The likelihood described above is performed 
for each sub season, for the union of all sub-seasons where a slope was
removed, and for the union of all data.  Similarly our analysis is
performed for each frequency channel independently and for the union
of all channels. As of the writing of this article the analysis is in
the final stages of checking the calibration; a full discussion of the
results will be presented \cite{farese03}.  

\section{Conclusion}
\CO was a first generation, scanning style CMB polarization
experiment.  While only setting an upper limit \CO has a number of
valuable lessons to teach us, or at least of which to remind us.

First in building a mount avoid the use of friction based drive
systems.  Less obviously, and quite triumphantly, \CO demonstrated
that the use of a Styrofoam secondary support in an on axis
configuration is sufficiently strong, stable and reliable.  One
must be prepared, however, to deal with dew condensation or other
possible effects of weather.

In terms of experimental design we have also had
emphasized two valuable points.  First, one must be particularly
careful about preventing, testing for, and removing,
systematic effects.  The more complicated nature and smaller signal of
polarized observations makes this even more demanding.  This leads to
the second point that in designing a scan strategy one should be sure
that one couples the mapping strategy well to the desired signal.  This is
to say that the eigenmodes measured by the experiment should be those
expected to have large signal.  Further one should be sure that the
probable systematic effects have eigenmodes as close to orthogonal to
the signal as possible.  The naive approximation of studying the noise
per pixel is not sufficient.

Finally, in analyzing the data two valuable insights were discovered.
The use of displaying the data in a variety of map formats is
tremendous, especially in that it allows the identification and removal
of systematic effects.  Also, one must be sure that an actual
likelihood curve, rather than a likelihood estimator, is calculated if
the expected signal to noise of the experiment is not much greater
than one.

\section{Acknowledgments}
We thank John Carlstrom for loaning us the required HEMT
amplifiers.  This research was supported by NSF grant AST-9802851. 

\bibliographystyle{plain} 
\bibliography{thesis}

\end{document}